# Discovery of Higher-Order Nodal Surface Semimetals

Huahui Qiu, Yuzeng Li, Qicheng Zhang, and Chunyin Qiu[*]

Key Laboratory of Artificial Micro- and Nano-Structures of Ministry of Education and School of Physics and Technology, Wuhan University, Wuhan 430072, China

[*] To whom correspondence should be addressed: cyqiu@whu.edu.cn

**Abstract.** The emergent higher-order topological insulators significantly deepen our understanding of topological physics. Recently, the study has been extended to topological semimetals featuring gapless bulk band nodes. To date, higher-order nodal point and line semimetals have been successfully realized in different physical platforms. However, the concept of higher-order nodal surface semimetals, the final frontier in this field, has yet to be proposed, let alone experimentally observed. Here, we report an ingenious design route for constructing this unprecedented higher-order topological phase. The three-dimensional model, layer-stacked with two-dimensional anisotropic Su-Schrieffer-Heeger lattice, exhibits appealing hinge arcs connecting the projected nodal surfaces. Experimentally, we realize this new topological phase in an acoustic metamaterial, and present unambiguous evidence for both the bulk nodal structure and hinge arc states, the two key manifestations of the higher-order nodal surface semimetal. Our findings can be extended to other classical systems such as photonic, elastic, and electric circuit systems, and open new possibilities for controlling waves.

*Introduction.* The discovery of topological insulators (TIs) has sparked immense interest in the field of topological phases of matter [1-3]. Recently, the study has been expanded to higher-order (HO) TIs, which goes far beyond the description of conventional topological band theory [4-11]. Markedly different from the conventional $n$-dimensional TIs that host nontrivial modes at $(n-1)$-dimensional boundaries, the HO TIs feature nontrivial responses at lower-dimensional sample boundaries, e.g., zero-dimensional (0D) corner modes in two-dimensional (2D) systems [4-23]. The HO phases and associated boundary manifestations are significantly enriched in three-dimensional (3D) systems [5,6,24-36]. In addition to those featuring 0D corner states, one can expect 3D HO TIs that exhibit topological response at their one-dimensional (1D) hinges [24-27]. The latter, also dubbed 3D second-order TI, is a direct 3D extension of the 2D HO TI by considering each $k_z$-slice as a 2D subsystem of HO topology, resulting in nontrivial hinge states across the projected 1D Brillouin zone (BZ) [Fig. 1(a)]. Similar pictures preserve for 3D gapless phases, yielding the so-called HO topological semimetals [28-36], in which nontrivial HO band topology can emerge in the gapped $k_z$-slices. Famous examples are HO nodal point (NP) semimetals (including HO Weyl [28-30] and Dirac [31-33] semimetals) and HO nodal line (NL) semimetals [34-36], which manifest nontrivial hinge arcs connecting the projected NPs [Fig. 1(b)] and NLs [Fig. 1(c)], respectively.

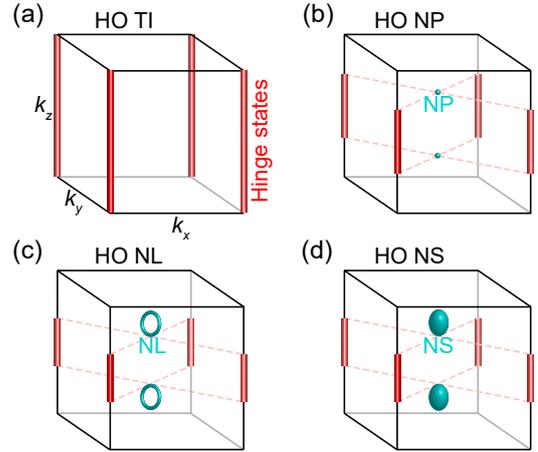

FIG. 1. Summary of 3D HO topological semimetals. In contrast to the 3D HO TI (a) with 1D hinge states over the projected BZ, the 3D HO NP (b), NL (c), and NS (d) semimetals feature hinge arcs connecting the projected nodal objects.

Experimentally, the HO NP semimetals have been realized in many distinctive physical platforms, including acoustic [37-41], photonic [42,43], and electric circuit [44] systems, etc. A variety of appealing phenomena have been observed, such as the coexisted surface and hinge arcs, and the bounded topological hinge states in the continuum. As for the HO NL semimetals, experimental implementations have only been reported in acoustics very recently [45,46], which demonstrate both the conventional drumhead surface states and HO hinge states. To the best of our knowledge, however, the HO topological semimetals with nodal surfaces (NSs) [47-49], the highest-dimensional nodal objects, have never been reported, either theoretically or experimentally. The significant challenge lies in identifying compatible symmetries that can protect the gapless NSs while simultaneously exhibiting HO topology in gapped $k_z$-slices.

In this Letter, we propose a simple model for constructing such unprecedented 3D HO NS semimetals. As depicted in Fig. 1(d), it hosts one pair of NSs in the 3D bulk BZ, and harbors nontrivial hinge states between the projected NSs. The latter acts as a key manifestation of the HO bulk-hinge correspondence. Theoretically, we start from a 2D anisotropic Su-Schrieffer-Heeger (ASSH) model that hosts distinctive NL and HO TI phases, alongside trivial insulators. By stacking the 2D ASSH model in the third direction, we achieve a 3D HO NS semimetal: the bulk NS structure, originating from the $k_z$-evolved NLs, is protected by the parity-time and chiral symmetries, while the HO topology in gapped $k_z$-slices is compatibly protected by the chiral symmetry. Experimentally, we design and fabricate a 3D acoustic metamaterial to emulate the tight-binding model. As a complete characterization of this long-desired topological phase, we not only detect the signal of the bulk NS structure through surface measurements, but also observe the appealing hinge states via reciprocal-space spectroscope and real-space visualization.

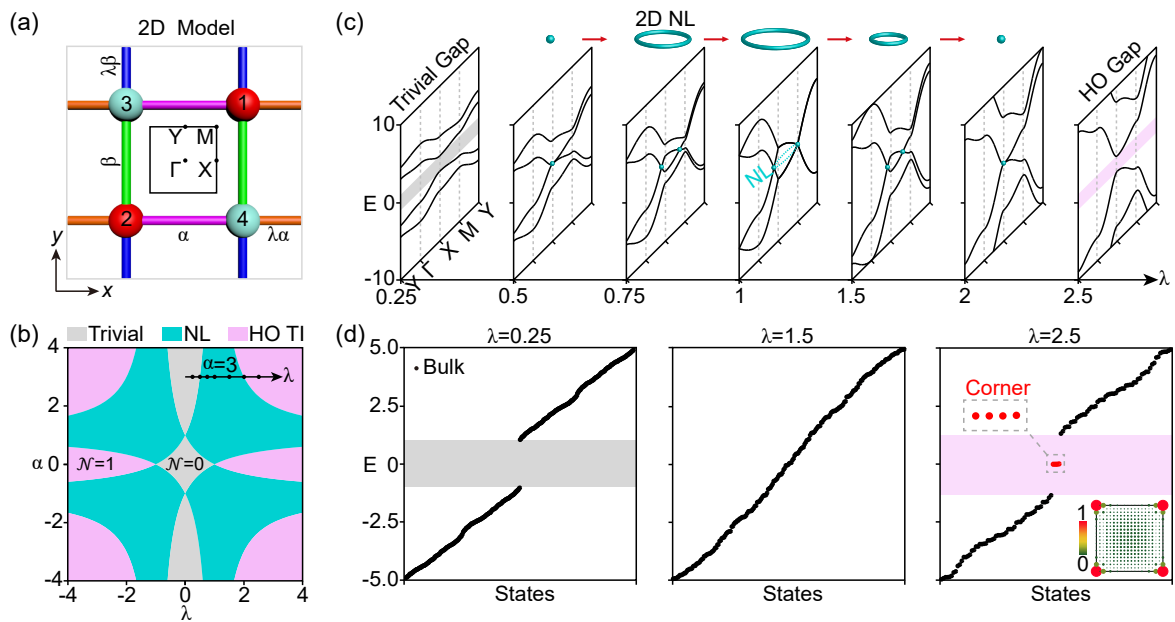

FIG. 2. 2D ASSH model. (a) Tight-binding model. The four orbitals coupled with *anisotropic* hoppings belong to two decoupled sublattices, as distinguished by the balls of different color. Inset: 2D BZ. (b) Phase diagram plotted for fixed $\beta = 1$. It includes a gapless NL semimetal phase, in addition to the gapped trivial insulator ($\mathcal{N} = 0$) and HO TI ($\mathcal{N} = 1$) phases. (c) Band structures for a sequence of systems with fixed $\alpha = 3$ but varied $\lambda$ [see the dots in (b)], which exhibit a ring-like 2D NL within $0.5 < \lambda < 2$. (d) Finite-size spectra for three specified $\lambda$ values. In contrast to the trivial insulator (left) and NL semimetal (middle) phases, the 2D HO TI (right) hosts four nontrivial zero-energy corner states (inset).

*Theoretical models.* As illustrated in Fig. 2(a), we begin with a 2D ASSH model [6,50,51]. It features intracell (intercell) hoppings $\alpha$ ($\lambda\alpha$) and $\beta$ ($\lambda\beta$) in the $x$ and $y$ directions, respectively. For brevity, here we consider an identical intercell/intracell hopping ratio $\lambda$ for both directions. The Bloch Hamiltonian of this model reads $\mathcal{H}_{2D}(\boldsymbol{k}) = \alpha(1 + \lambda \cos k_x)\Gamma_{10} - \lambda\alpha \sin k_x \Gamma_{23} + \beta(1 + \lambda \cos k_y)\Gamma_{11} - \lambda\beta \sin k_y \Gamma_{12}$, where $\Gamma_{ij} = \tau_i \otimes \sigma_j$ with $i,j \in [0,3]$, and $\tau_i$ and $\sigma_j$ are Pauli matrices for the degree of freedom within a unit cell. In addition to $C_{2v}$ and time-reversal symmetries (thus parity-time symmetry in 2D, automatically), the 2D lattice enjoys a sublattice symmetry (also known as chiral symmetry). Instead of using an *isotropic* model ($\alpha = \beta$) [50,52,53], where the HO topology has been unveiled for the first and third bandgaps, here we consider the *anisotropic* case ($\alpha \neq \beta$) and focus on the tunable second bandgap around zero energy. This gives three different 2D topological phases [Fig. 2(b)], i.e., trivial insulator, NL semimetal, and HO TI, which is crucial for constructing our 3D HO NS semimetal later. Note that the HO topology in this middle gap cannot be identified by a 2D polarization [50,52,53] or $C_2$ invariant [54] as usual. It is distinguished by a new integer topological invariant ($\mathcal{N}$) defined in real space, called quadrupole chiral number [55-58], where $|\mathcal{N}|$ characterizes the number of states localized at each corner (Supplemental Material [59]). Figure 2(c) exemplifies the band structures for a series of systems with fixed $\alpha = 3$ but different $\lambda$. It shows that the middle gap appearing for small $\lambda$ vanishes at $\lambda = 0.5$, and the band node grows into a ring-like NL in the presence of the chiral and parity-time symmetries (Supplemental Material [59]); the zero-energy NL shrinks to a point at $\lambda = 2$, and then gaps out to form a nontrivial 2D HO TI for $\lambda > 2$. These facts are further confirmed by the corner spectra for three representative $\lambda$ values [Fig. 2(d)], which display clearly a trivial gap, a gap closure, and a HO gap with four degenerate zero-energy corner states (i.e., one state per corner as predicted by $\mathcal{N} = 1$).

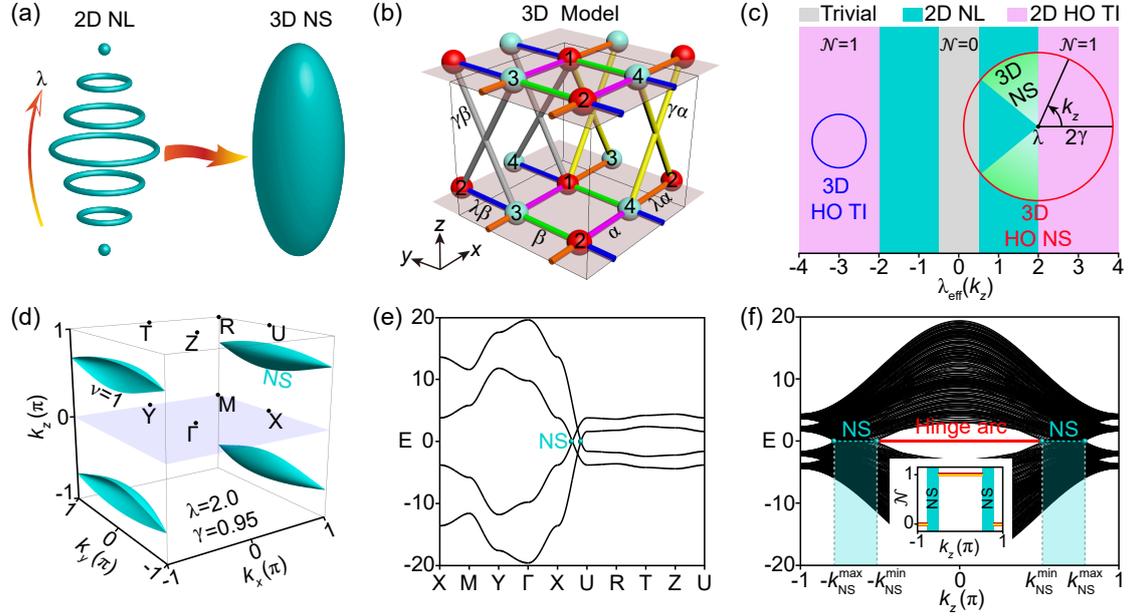

FIG. 3. Tight-binding model for constructing 3D HO NS semimetals. (a) Schematic of sweeping 2D NLs to form a 3D NS. (b) 3D lattice stacked by the 2D ASSH model. (c) Conceiving 3D topologically distinct phases from the effective hopping ratio $\lambda_{\text{eff}}(k_z) = \lambda + 2\gamma \cos k_z$. Again, we consider $\alpha = 3$ and $\beta = 1$. The blue and red circles sketch possible $k_z$-slice evolutions to generate a 3D HO TI and a 3D HO NS, respectively. (d) Visualization of the NS structure for the specified $\lambda$ and $\gamma$. Each rugby-like NS carries a nontrivial $\mathbb{Z}_2$ charge $\nu = 1$. (e) Bulk band structure along high-symmetry lines. (f) Hinge-projected spectrum, which features zero-energy hinge arcs (red line) connecting the projected NSs. Inset: Quadrupole chiral number $\mathcal{N}$ plotted as a function of $k_z$.

Figure 2(c) suggests that one may conceive a 3D NS by sweeping 2D NLs continuously with the hopping ratio $\lambda$ [Fig. 3(a)]. To do this, we stack the 2D ASSH model in the $z$ direction via introducing interlayer couplings $\gamma\alpha$ and $\gamma\beta$ with $\gamma > 0$ [Fig. 3(b)]. The 3D system can be readily modeled by the Hamiltonian $\mathcal{H}_{\text{2D}}$, where $\lambda$ is replaced by $\lambda_{\text{eff}}(k_z) = \lambda + 2\gamma \cos k_z$. As $k_z$ varies from $-\pi$ to $\pi$, the effective parameter $\lambda_{\text{eff}}$ experiences a circle of radius $2\gamma$ around the origin specified with $\lambda$, which enables topologically distinct 3D phases according to the evolution scenarios of $k_z$-slices (Supplemental Material [59]). As exemplified by the blue circle in Fig. 3(c), when the origin falls in the nontrivial phase ($\mathcal{N} = 1$) and the radius is sufficiently small, any $k_z$-slice realizes the 2D HO TI phase, and the stacked system constitutes a 3D HO TI featuring perfectly flat zero-energy hinge states due to the chiral symmetry. By contrast, when the parametric loop (red circle) traverses the boundaries of the NL phase, one time-reversal pair of NSs emerge and the system forms a HO NS semimetal. To see this, we consider a concrete system with $\lambda = 2.0$ and $\gamma = 0.95$. Figure 3(e) shows its bulk dispersion, where the two zero-energy linear crossing points signify the NS degeneracy along the momentum path X-U. More intuitively, we visualize the global nodal structure in 3D BZ [Fig. 3(d)], according to the analytical expression derived for zero-energy degeneracy, $|\alpha(1 + \lambda_{\text{eff}} e^{ik_x})| = |\beta(1 + \lambda_{\text{eff}} e^{ik_y})|$. The NS is protected by a nontrivial $\mathbb{Z}_2$ charge $\nu = 1$ in the presence of chiral and parity-time symmetries [47] (Supplemental Material [59]). Figure 3(f) presents the hinge spectrum in the $z$ direction. It shows clearly four degenerate 1D hinge arcs (i.e., one arc per hinge) that link the projected NSs within $k_{\text{NS}}^{\min} < |k_z| < k_{\text{NS}}^{\max}$. The result is consistent with the theoretical prediction of the $k_z$-distributed quadrupole chiral number (see inset), $\mathcal{N} = 0$ for $|k_z| > |k_{\text{NS}}^{\max}|$ while $\mathcal{N} = 1$ for $|k_z| < |k_{\text{NS}}^{\min}|$. Note that although the topological invariant can also be defined for the gapped $k_x$ and $k_y$ slices, our calculations demonstrate $\mathcal{N} = 0$ and thus no zero-energy states emerge at those $x$- and $y$-directed hinges (Supplemental Material [59]).

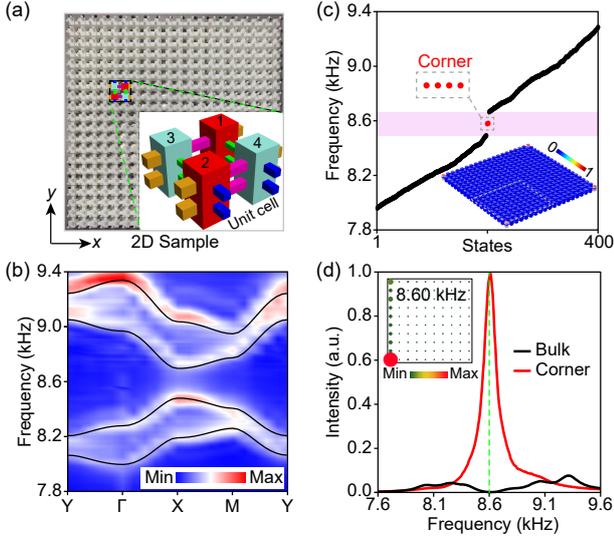

FIG. 4. Acoustic demonstration of the HO topology in 2D ASSH model. (a) Experimental sample of $10 \times 10$ unit cells. Inset: Unit cell geometry. (b) Simulated (lines) and measured (color) bulk dispersions. (c) Frequency spectrum simulated for the finite-size sample, which shows four degenerate corner states pinned to the middle of bulk gap. Inset: Pressure intensity profile of the corner mode, where the gray line sketches 1/4 sample. (d) Intensity spectra measured at bulk and corner sites. Inset: Intensity pattern extracted for the peak frequency of the corner spectrum, demonstrated with 1/4 sample for clarity.

*Acoustic realizations.* Before experimentally demonstrating the 3D HO NS semimetal, we verify the 2D HO topology of the middle gap in the ASSH model, which has yet to be confirmed in any experiment. Figure 4(a) shows our experimental sample fabricated by 3D printing technology. Each unit cell comprises four identical cuboid air cavities interconnected by narrow tubes (inset). Physically, the cavity resonators emulate atomic orbitals while the narrow tubes mimic the hoppings between them. With structure details supplied in Supplemental Material [59], this acoustic lattice captures well the tight-binding model with an onsite energy 8.60 kHz, intracell hoppings $\alpha = -117.8$ Hz and $\beta = -38.0$ Hz, and a hopping ratio $\lambda = 3.3$. For this anisotropic 2D system, our numerical band structure shows a sizeable complete bulk gap [Fig. 4(b), lines], within which four mid-gap corner states emerge as expected in the finite-size spectrum [Fig. 4(c)]. Experimentally, we placed a point-like sound source in the middle of the sample and scanned the pressure distribution over it (Supplemental Material [59]). After a 2D Fourier transformation, we mapped out the bulk dispersion of the sample, which reproduces the numerical one excellently [Fig. 4(b)]. Furthermore, we measured the local spectral responses for both the bulk and corner sites. As shown in Fig. 4(d), the corner spectrum exhibits a prominent peak around 8.60 kHz, in stark contrast to the dip bulk response to the middle gap. This distinctive spectral signature, along with the corner-localized sound profile at the peak frequency (inset), conclusively evidences the HO topology of our 2D acoustic lattice.

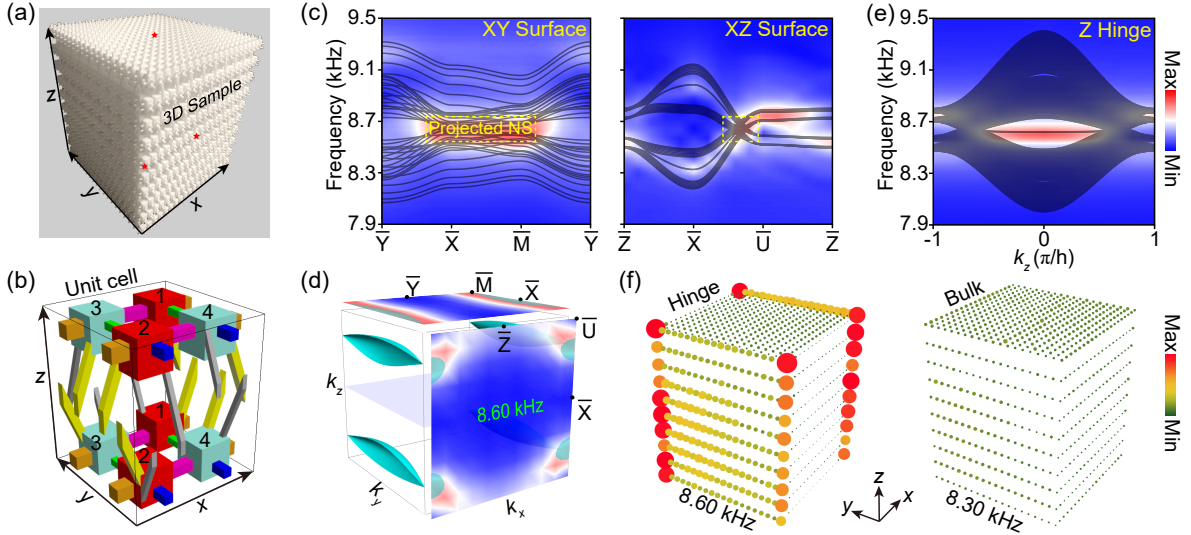

FIG. 5. Experimental evidence for our 3D acoustic HO NS semimetal. (a) Experimental sample. The red stars label the sound sources positioned for measuring the surface and hinge spectra. (b) Unit cell structure. The yellow and gray tubes specify the interlayer couplings. (c) Measured (color) and simulated (lines) surface spectra for the XY and XZ surfaces. The dashed yellow boxes signify the projected NS structure. (d) Equifrequency map extracted at 8.60 kHz (color), which capture the NS projections (shallow blue) to different surface BZs. (e) Hinge spectra measured and simulated along the $k_z$ direction. (f) Pressure distributions measured at 8.60 and 8.30 kHz, clearly showing the hinge (left) and bulk (right) states, respectively.

Now we turn to the acoustic realization of the 3D HO NS semimetal. According to the tight-binding model in Fig. 3(b), we elaborately devised an acoustic sample of $10 \times 10 \times 10$ unit cells [Figs. 5(a)], whose unit cell structure is sketched in Fig. 5(b). This acoustic lattice emulates the tight-binding model with $\alpha = -105.0$ Hz, $\beta = -35.0$ Hz, $\lambda = 2.0$, and $\gamma = 0.95$ (Supplemental Material [59]). To identify the nodal structure in bulk bands, acoustic measurements were implemented on the XY and XZ surfaces, where the sound source was positioned in the middle of the corresponding surface. (Direct bulk measurements are inaccessible due to the great challenge in locating the sound source and detector inside the sample.) Figure 5(c) presents the experimentally measured frequency spectra. As expected, the XY-surface spectrum shows a strong signal at ~8.60 kHz along the $\overline{\text{XM}}$ direction, due to the high density of states at the NS frequency; meanwhile, the YZ-surface spectrum exhibits a bright spot in the momentum path $\overline{\text{XU}}$, which signifies the projected NSs on the XZ surface. The bulk nodal structure can be further confirmed by the iso-frequency maps extracted at the NS frequency [Fig. 5(d)], where the bright regions indicate the projected NSs on the two surface BZs. All above experimental data agree reasonably well with the model predictions.

To verify the HO topology of the $k_z$-slices between the pair of NSs, the other vital hallmark of the 3D HO NS semimetal, we relocated the sound source to the middle of the Z hinge and scanned the pressure response along it. Figure 5(e) presents the associated hinge spectrum in momentum space. Unambiguously, the bright spot evidence the hinge arc between the projected NSs. To visualize the hinge states more clearly, we measured the site-resolved local response, in which the sound source and probe were placed in the same cavities. Figure 5(f) presents the intensity profiles extracted for the NS frequency (8.60 kHz) and a frequency away from it (8.30 kHz). It shows a strong sound localization at 8.60 kHz along the *z*-directed hinge sites, in stark contrast to the evenly distributed bulk states at 8.30 kHz. Potential applications could be anticipated for the such highly localized HO hinge modes, e.g., acoustic sensing and energy trapping.

*Conclusions*. We have proposed the first 3D HO NS semimetal and demonstrated it in acoustics unambiguously. Experimentally, we have not only characterized the unique nodal structure from the projected bulk states, but also identified the HO topology from the hinge responses. All experimental data capture well our theoretical predictions. Note that our 3D model can also be applied to realize other HO topological phases, such as 3D HO NL semimetals via breaking inversion symmetry (Supplemental Material [59]). There are still more intriguing topological effects waiting to be explored in 3D HO NS semimetals. For instance, introducing extra non-Hermiticity may bring about new fascinating phenomena, such as hinge skin modes [60,61].

**Acknowledgements**
This project is supported by the National Natural Science Foundation of China (Grant No. 12374418, 11890701, and 12104346), and the Young Top-Notch Talent for Ten Thousand Talent Program (2019-2022).